\documentclass[preprint,aps,amsmath,superscriptaddress,nofootinbib,tightenlines]{revtex4}
\usepackage{graphicx}
\usepackage{bm}
\usepackage{epsfig}
\usepackage{ulem}

\renewcommand{\emph}[1]{{\it #1}}

\def\nslash{n\!\!\!\slash}
\def\bnslash{\bar n\!\!\!\slash}

\newcommand{\nn}{\nonumber} 
\newcommand{\bn}{{\bar n}}
\newcommand{\bea}{\begin{eqnarray}}
\newcommand{\eea}{\end{eqnarray}}
\newcommand{\mcdot}{\!\cdot\!}

\newcommand{\be}{\begin{equation}}
\newcommand{\ee}{\end{equation}}
\def \ba  {\begin{eqnarray}}
\def \ea  {\end{eqnarray}}
\newcommand{\vect}[1]{\mathbf{#1}}
\newcommand{\abs}[1]{\left\lvert #1\right\rvert}
\newcommand{\bra}[1]{\left\langle #1\right\rvert}
\newcommand{\ket}[1]{\left\lvert #1\right\rangle}
\newcommand{\Lqcd}{\Lambda_{\text{QCD}}}
\newcommand{\e}{\mathrm{e}}
\newcommand{\boldeps}{\mbox{\boldmath$\varepsilon$}}
\newcommand{\eq}[1]{Eq.~\eqref{#1}}
\newcommand{\eqs}[2]{Eqs.~\eqref{#1} and \eqref{#2}}

\DeclareMathOperator{\Tr}{Tr}

\newcommand{\tL}{t_L}

\begin{document}


\preprint{UCB-PTH-08/02}
\preprint{YITP-SB-08-02}

\title{Factorization of $e^+e^-$ Event Shape Distributions with Hadronic Final States in Soft Collinear Effective Theory}

\author{Christian W.~Bauer\footnote{Electronic address: cwbauer@lbl.gov}}
\affiliation{Ernest Orlando Lawrence Berkeley National Laboratory and
University of California, Berkeley, CA 94720, USA}

\author{Sean Fleming\footnote{Electronic address: fleming@physics.arizona.edu}}
  \affiliation{University of Arizona, Tucson, AZ 85721, USA}
  
  \author{Christopher Lee\footnote{Electronic address: clee@berkeley.edu}}
  \affiliation{Ernest Orlando Lawrence Berkeley National Laboratory and
University of California, Berkeley, CA 94720, USA}

\author{George Sterman\footnote{Electronic address: sterman@insti.physics.sunysb.edu}}
\affiliation{C.N.\ Yang Institute for Theoretical Physics,\\ Stony Brook University,
Stony Brook, NY 11794-3840, USA}

\date{\today\\ \vspace{1cm} }


\begin{abstract}
We present a new analysis of two-jet event shape distributions in soft collinear 
effective theory.  Extending previous results, we observe that a large class of such distributions can be expressed in terms of vacuum matrix elements of operators in the effective theory. We match these matrix elements  to the full theory in the two-jet limit without 
assuming factorization of the complete set of hadronic final states into independent 
sums over partonic collinear and soft states. We also briefly discuss the relationship of 
this approach to diagrammatic factorization in the full theory.
\end{abstract}

\maketitle

\newpage

\section{Introduction}

Hadronic jets provide a window into the fundamental workings of quantum chromodynamics, since they contain within themselves the signatures of QCD at both weak and strong coupling. That hadronic final states in high-energy collisions are jet-like at all reflects the underlying parton-level perturbative interactions at weak coupling $\alpha_s\ll 1$, while the evolution of individual partons into the final-state hadrons we detect  depends on the transition to nonperturbative dynamics at strong coupling. Jets therefore test our understanding of QCD over a wide range of scales.

At the same time that jets provide insights into QCD, they are key elements in signatures for new physics beyond the Standard Model. Unfortunately, the challenge of calculating QCD background events in the hadron collider environment of the Tevatron or the LHC remains formidable. To develop incrementally our ability to describe the structure and evolution of jets in these environments,  it is useful to continue to develop the analysis of jets produced in the relatively cleaner environment of $e^+ e^-$ collisions. 

On the one hand, one can study observables that depend on specifying the actual number of jets in the final state by defining a jet algorithm, which sets criteria for what constitutes a jet. On the other hand, one can extract much useful information about the structure of the final state from simpler observables called \emph{event shapes}, which do not depend on a jet algorithm, but are simple functions of the momenta of all the particles in the final state. \emph{Two-jet event shapes} $e$ are those whose distribution near $e= 0$ is dominated by events with two nearly back-to-back collimated jets of particles. This  is the case for $e^+ e^-$ annihilation at high energy, and there is no need to specify precisely what constitutes a jet. Some familiar examples are thrust $\tau = 1-T$~\cite{Brandt:1964sa,Farhi:1977sg}, jet broadening $B$~\cite{Catani:1992jc}, jet masses~\cite{Chandramohan:1980ry,Clavelli:1979md,Clavelli:1981yh}, and the $C$-parameter~\cite{Ellis:1980wv}. A generic class of event shapes for events $e^+ e^-\rightarrow X$ at center-of-mass energy $Q$ may be defined by
\begin{equation}
\label{eventshpdef1}
e(X) = \frac{1}{Q}\sum_{i\in X}f_e(\eta_i)\abs{\vect{p}_i^{\rm T}},
\end{equation}
where $\eta_i$ and $\vect{p}_i^{\rm T}$ are the rapidity and transverse momentum of final state particle $i$ with respect to the thrust axis of the event. The function $f_e$ is sufficiently well behaved to guarantee infrared safety. Choices of $f_e$ that give familiar event shapes are
\begin{equation}
f_\tau(\eta) = \e^{-\abs{\eta}},\qquad f_B(\eta) = 1,\qquad f_C(\eta) = \frac{3}{\cosh\eta}\,.
\end{equation}
A recently-introduced class of event shapes $\tau_a$, the \emph{angularities}~\cite{Berger:2003iw,Berger:2003pk}, for which
\begin{equation}
\label{oureventshp}
f_a(\eta) = \e^{-\abs{\eta}(1-a)},
\end{equation}
generalizes thrust and jet broadening. The thrust corresponds to $a=0$, and the broadening to $a=1$. Any $a$ with $-\infty<a<2$ defines an infrared-safe observable. Studying how the behavior of the angularity distributions vary with $a$ provides insight into the intrinsic structure of jets \cite{Berger:2003pk,Berger:2004xf}.

In the endpoint region of two-jet event shape distributions, fixed-order perturbation theory alone is insufficient to make accurate predictions. This region is dominated by perturbative logarithms of $e$ and by nonperturbative  power corrections. A tool to separate the perturbative and nonperturbative contributions and set up a resummation of the logarithms in these distributions is \emph{factorization}, which separates the effects of the hard scattering, jet evolution, and hadronization occurring at different length and energy scales. An observable factorizes if it can be calculated as the convolution of perturbatively-calculable functions and nonperturbative functions, which are typically universal in the sense that the same nonperturbative function contributes to multiple processes. A two-jet event shape distribution that factorizes takes the form
\begin{equation}
\frac{1}{\sigma_0}\frac{d\sigma}{de} = H(\mu) \int de_1\,de_{2}\,de_s\,\delta(e-e_1-e_{2}-e_s) J_1(e_1;\mu) J_{2}(e_{2};\mu) S(e_s;\mu),
\label{factorizationtheorem}
\end{equation}
where $H(\mu)$ is a perturbatively-calculable hard function, and $J_i$ and $S$ are 
jet and soft functions respectively.  
This form summarizes the leading-power behavior, $1/e\ \times$ logarithms of $e$, of the cross-section. Corrections are suppressed by powers of $e$.

It is important to emphasize that the precise definitions of the jet and soft functions
are somewhat arbitrary, and differ from treatment to treatment.
Analyzing both of these functions perturbatively, one finds logarithmic dependence on the
event shape variables $e_i$ and $e_s$. This dependence can be resummed,
from which we can generate the resummed dependence of the cross-section on $e$.
The original resummations of thrust and related event shape variables \cite{Catani:1991kz} used the feature that the entire next-to-leading logarithmic behavior
is naturally absorbed into the jet functions.  These discussions do not need to
include a soft function.    The general structure of \eq{factorizationtheorem},
including both soft and jet functions, and its relationship to
Sudakov resummation at arbitrary logarithmic level was explored
in a general context in Ref.~\cite{Contopanagos:1996nh}. Perturbative applications to jet shapes
were described in more detail, with explicit constructions for the jet
and soft functions in Ref.~\cite{Berger:2003iw}.
These treatments assume that perturbation theory provides a leading-power
description of infrared-safe observables in the manner of an
asymptotic series, with power corrections whose behavior can
be inferred from the structure of the perturbative series at
high orders, and/or the running of the coupling \cite{Beneke:2000kc}.
Both the jet and soft functions also generally receive power corrections
in this manner,
and nonperturbative corrections to event shapes
based on both fixed-order and resummed cross sections have been widely discussed,
and reviewed in \cite{Dasgupta:2003iq}.

Recently the factorization of jet cross sections and event shape distributions near the two-jet kinematic endpoint has been
revisited in the language of soft collinear effective theory (SCET)~\cite{Bauer:2000ew,Bauer:2000yr,Bauer:2001ct,Bauer:2001yt}. This approach provides an elegant way, developed in \cite{Bauer:2002ie,Bauer:2002nz,Bauer:2003di,Lee:2006nr}, to reproduce the results of the traditional QCD factorization at leading power, and provides a framework to organize perturbative resummations~\cite{Bauer:2006mk,Bauer:2006qp,Trott:2006bk,Schwartz:2007ib} and nonperturbative power corrections~\cite{Lee:2006nr,Hoang:2007vb}. These analyses of massless jets have also been extended to the case of massive jets~\cite{Fleming:2007qr,Fleming:2007xt}.
The effective field theory also provides
a systematic framework to move beyond
leading-power
results.  In SCET, the factorization in Eq.~(\ref{factorizationtheorem}) follows from the usual separation of short- and long-distance physics in effective field theories, as well as from the decoupling of interactions between collinear and soft fields in the leading-order SCET Lagrangian. The hard function $H$ is the square of a Wilson coefficient from the matching between QCD currents and SCET operators, the jet functions $J_i$ are matrix elements of operators built of collinear fields, and the soft function $S$ is the matrix element of an operator built from ultrasoft (usoft) fields.

One goal of the SCET formulations of factorization in \cite{Bauer:2002ie,Bauer:2002nz,Bauer:2003di,Lee:2006nr,Fleming:2007qr} has been
to use effective theory methods to improve our insight into nonperturbative corrections
by  analyzing matrix elements directly in terms of hadronic states.   As we shall see, the jet cross sections that we study below can be reformulated in terms of hadronic matrix elements, involving nonlocal products of currents and the energy-momentum tensor \cite{Sveshnikov:1995vi,Korchemsky:1997sy}.
There is an important subtlety, however, in the application of effective field theory
methods to semi-inclusive sums over asymptotic states.
This is the assumption that
hadronic
final states $X$ can be factorized
consistently as
\begin{equation}
\ket{X} = \ket{X_c}\ket{X_{\rm soft}}
\label{phase_space_equation}
\end{equation}
where $X_c$ and $X_{\rm soft}$ are collinear and soft
final  states, which carry non-singlet color in general.
Indeed, it is inconsistent to impose a color-singlet
condition on the collinear and soft states individually
\cite{Bauer:2003di}.  Furthermore,
the relation \eq{phase_space_equation} must
be associated with power corrections  that in principle
depend on the observable to which we intend to apply it.  Thus, corrections
to \eq{phase_space_equation} are not
well-defined without additional input.
Similar limitations apply to an appeal
to parton-hadron duality to justify \eq{phase_space_equation},
since this principle does not come with a systematic
method of estimating corrections.
Finally, if we were to take both sides of the relation literally,
it would have to hold at infinite times, which would violate
the spirit of matching the effective to full theory
at a perturbative scale,  and only then evolving
to nonperturbative scales.    These considerations
motivate our analysis below.

In this paper, we show that the assumption Eq.~(\ref{phase_space_equation}) is not necessary to prove the factorization Eq.~(\ref{factorizationtheorem}). We show that the phase space
delta function $\delta(e-e(X))$, which restricts the final states in $d\sigma/de$ to the events with $e(X) = e$, can be expressed as an operator $\delta(e-\hat e)$ acting on the final state $X$. The operator $\hat e$, which can be constructed in quantum field theory,  depends on an energy flow operator as well as an operator $\hat t$, which picks out the thrust axis of a final state $X$. Using these ingredients to express $\delta(e-e(X))$ as an operator, we remove its dependence on the final state $X$ and so are able to sum over the complete set of states before it ever becomes necessary to assume the factorization of states, Eq.~(\ref{phase_space_equation}).

In the next section, we demonstrate the factorization of the event shape distribution $d\sigma/de$ in SCET, making use of the event shape operator $\hat e$. In Section~\ref{sec:defofoperators}
we construct
the energy flow operator that enters in the construction of $\hat e$ and discuss the determination within SCET of the thrust axis on which $\hat e$ depends.
Section~\ref{sec:conclusions} contains our conclusions.

\section{Factorization proof}
\label{sec:twojet}

\subsection{Event shape distributions in QCD}

We start with the distribution in the event shape $e$ in $e^+e^-\rightarrow {\rm hadrons}$, which is given in full QCD by
\begin{equation}
\frac{d\sigma}{de} = \frac{1}{2Q^2}\sum_X\abs{\mathcal{M}(e^+e^-\rightarrow X)}^2 (2\pi)^4\delta^4(q-p_X)\delta(e-e(X))\,,
\label{QCD1}
\end{equation}
where $q$ is the total incoming momentum, with $q^2 \equiv Q^2$. Here $X$ labels the hadronic final state, and $e(X)$ denotes the value of the event shape variable for a given final state $X$. To leading order in the electroweak couplings, the process $e^+ e^- \to X$ is mediated by the partonic $s$-channel transition $e^+ e^- \to q \bar q$, with an intermediate photon or $Z$ boson. The leptonic part of this partonic transition matrix element can be calculated, and one finds
\be
\abs{\mathcal{M}(e^+e^-\rightarrow X)}^2 = \sum_{i=V,A}L^i_{\mu\nu} \bra{0}j_i^{\mu\dag} \ket{X}\bra{X}j_i^\nu\ket{0}\,,
\label{amplitudesquared}
\ee
where we have defined the vector and axial currents,
\be
\label{QCDcurrent}
j_i^\mu = \bar q_f^a \Gamma_i^\mu q_f^a\,,
\ee
with $\Gamma_V^\mu = \gamma^\mu$ and $\Gamma_A^\mu = \gamma^\mu\gamma^5$.  The leptonic tensor is given by
\begin{subequations}
\begin{align}
L^V_{\mu\nu} &= -\frac{e^4}{3Q^2}\left(g_{\mu\nu} - \frac{q_\mu q_\nu}{Q^2}\right)\left[Q_f^2 - \frac{2Q^2 v_e v_f Q_f}{Q^2 - M_Z^2} + \frac{Q^4(v_e^2+a_e^2)v_f^2}{(Q^2-M_Z^2)^2}\right] \\
L^A_{\mu\nu} &= -\frac{e^4}{3Q^2}\left(g_{\mu\nu} - \frac{q_\mu q_\nu}{Q^2}\right)\frac{Q^4(v_e^2+a_e^2)a_f^2}{(Q^2-M_Z^2)^2}\,, 
\end{align}
\end{subequations}
where fermion $f$ has electric charge $Q_f$ in units of $e$, and vector and axial charges $v_f,a_f$ given by 
\begin{equation}
v_f = \frac{1}{2\sin\theta_W\cos\theta_W}(T^3_f - 2Q_f\sin^2\theta_W),\quad a_f = \frac{1}{2\sin\theta_W\cos\theta_W}T^3_f.
\end{equation}
In Eq.~(\ref{QCDcurrent}) a sum over colors $a$ and flavors $f$ is understood.

Writing the four-momentum conserving delta function as the integral of an exponential, and using the dependence on $p_X$ in the exponential to translate one of the two currents to the position $x$ we can write the distribution as
\be
\frac{d\sigma}{de} = \frac{1}{2Q^2}\sum_X\int d^4 x\,\e^{iq\cdot x}\sum_{i=V,A}L^i_{\mu\nu}\bra{0} j_i^{\mu\dag}(x)\ket{X}\bra{X}j_i^\nu(0)\ket{0}\delta(e - e(X))\,,
\label{QCDdist}
\ee

\subsection{Eliminating the dependence on the final state}

The delta function $\delta(e - e(X))$ restricts the sum over final states to those states giving the same value $e$ of the observable event shape. This means that we cannot perform the sum over the complete set of final states. However, as we will now show, it is possible to write the event shape $e(X)$ as the eigenvalue of an operator acting on the final state $X$. This can be achieved using the definition of the transverse energy flow operator  $\mathcal{E}_T(\eta)$, which was introduced in~\cite{Lee:2006nr}. Its action on a hadronic state $X $ is given by
\be
\mathcal{E}_T(\eta) \ket{X} = \sum_{i\in X}\abs{\vect{p}_i^T}\delta(\eta-\eta_i)\ket{X},
\label{ETdef}
\ee
where $\vect{p}_i^T$ is the transverse momentum of the $i$th particle with respect to the thrust axis, and $\eta_i$ is the rapidity of the $i$th particle. The thrust axis is defined to be the unit vector $\vect{t}$ which maximizes the sum $\sum_i \abs{\vect{p}_i\cdot \vect{{t}}}$.  In the event shapes of \eq{eventshpdef1}, rapidities and transverse momenta are measured with respect to this axis. Thus implicit in the action of $\mathcal{E}_T(\eta)$ on $\ket{X}$ is the determination of this thrust axis $\vect{t}(X)$. Using the energy flow operator we define an operator $\hat e$, which returns the value of the event shape for a given state $X$,
\be
\hat e \ket{X} \equiv e(X) \ket{X} = \frac{1}{Q} \int_{-\infty}^\infty \!\!d\eta \, f_e(\eta) \mathcal{E}_T(\eta;\hat t) \ket{X}\,,
\label{ehatdef}
\ee
where $\hat t$ is an operator that returns the value of the thrust axis $\vect{t}(X)$ when acting on the final state $X$, and we have denoted explicitly the dependence of $\mathcal{E}_T(\eta)$ on this axis.
Although the present argument does not rely on explicitly constructing $\hat t$, it is nevertheless possible to do so, as we show in Sec.~\ref{sec:thrustQCD}.
In Sec.~\ref{sec:thrustSCET} we argue that in SCET we can choose the thrust axis to be in the jet direction $\vect n$ appearing in the two-jet current, so that no $\hat t$ operator need act on the final state at all.

Using the thrust axis and event shape operators $\hat t$ and $\hat e$, we can remove all dependence on the final state in the factor $\delta(e-e(X))$ in \eq{QCDdist} and can therefore perform the sum over the complete set of final states. 
This gives
\be
\frac{d\sigma}{de} = \frac{1}{2Q^2}\int d^4 x\,\e^{iq\cdot x}\sum_{i=V,A} L^i_{\mu\nu}\bra{0} j_i^{\mu\dag}(x)\delta(e - \hat e)j_i^\nu(0)\ket{0}\,.
\label{QCD2}
\ee
The expression above involves a delta function of the operator $\hat{e}$, which requires further comment. Heuristically this delta function is a way of treating the factorization of all moments of $\hat{e}$ at the same time. To see this we first define the delta function operator $\delta(e - \hat e)$ through a Taylor series expansion in $\hat{e}$: 
\begin{equation}
\delta(e - \hat e) = \delta(e) + \hat{e} \, \delta^{(1)}(e) +  \cdots + \frac{\hat{e}^n}{n!} \, \delta^{(n)}(e) + \cdots \,.
\end{equation}
{}From this expression it is clear that the $n$th term in the series is the $n$th moment of the event shape distribution. Thus if we integrate Eq.~(\ref{QCD2}) against $e^n$ the delta function operator on the right picks out the $n$th moment of the event shape distribution.

In order to factorize this matrix element, we need to match the full theory currents onto operators in SCET, and to construct explicitly the operator $\hat e$ in SCET. The operator $\mathcal{E}_T(\eta)$ is related to the energy-momentum tensor by \cite{Sveshnikov:1995vi,Korchemsky:1997sy}
\begin{equation}
\mathcal{E}_T(\eta) = \frac{1}{\cosh^3\eta}\int_0^{2\pi} d\phi\lim_{R\rightarrow\infty} R^2\int_0^\infty dt\,\hat n_i T_{0i}(t,R\hat n)\,.
\label{ETfromT0i}
\end{equation}
In Sec.~\ref{sec:EMtensor}, we will prove \eq{ETfromT0i} using the energy-momentum tensor $T_{\mu\nu}$ written in terms of fields corresponding to the hadrons in the state $X$. In the proof of factorization below, we will instead use its presumably equivalent form in terms of quark and gluon fields in QCD and SCET.  We are free to use either form as an operator is independent of its representation.

\subsection{Matching onto SCET}

We now match the currents and energy flow operator onto SCET. The discussion in this section is purposely kept brief, and for more details of the techniques used and for definitions of our notation we refer the reader to Refs.~\cite{Lee:2006nr,Fleming:2007qr,Bauer:2003di}. To reproduce the endpoint region of the two-jet event shape distribution, we match the QCD currents $j_i^\mu$ onto SCET operators containing fields in only two collinear directions:
\begin{eqnarray}
j_i^\mu(x) &=& \sum_{\vect{n}_1,\vect{n}_2}\sum_{\tilde p_1, \tilde p_2}C_{n_1 n_2}(\tilde p_1,\tilde p_2;\mu) {\cal O}_{n_1 n_2}(x; \tilde p_1, \tilde p_2)\,.
\label{SCETcurrentO2}
\end{eqnarray}
The operator ${\cal O}$ can depend on the label directions $n_1$ and $n_2$, as well as the label momenta $\tilde p_1,\tilde p_2$.  Recall that, in SCET, collinear momenta $p_c^\mu = \tilde p^\mu + k^\mu$ are divided into a large label piece, $\tilde p^\mu = (\bn\cdot \tilde p)n^\mu/2 + \tilde p_\perp^\mu$, and a residual piece, $k^\mu$, where $\bn\cdot\tilde p$ is $\mathcal{O}(Q)$, $\tilde p_\perp$ is $\mathcal{O}(Q\lambda)$, and $k$ is $\mathcal{O}(Q\lambda^2)$. The residual momenta are the same size as ultrasoft momenta, $k_{\text{us}}$, of $\mathcal{O}(Q\lambda^2)$.  The small parameter $\lambda$ is of order $\sqrt{\Lqcd/Q}$. 
After the Bauer-Pirjol-Stewart (BPS) field redefinition with usoft Wilson lines \cite{Bauer:2001yt}, the current in SCET is given by
\be
 {\cal O}_{n_1 n_2}(x; \tilde p_1, \tilde p_2)= e^{i(\tilde p_1 - \tilde p_2)\cdot x}\bar\chi_{n_1,p_1}(x)Y_{n_1}(x)\Gamma_i^\mu\overline Y_{n_2}(x)\chi_{n_2,p_2}(x)\,,
 \label{O2def}
 \ee
where $\Gamma_V^\mu = \gamma_\perp^\mu$ and $\Gamma_A^\mu = \gamma_\perp^\mu\gamma_5$. 
In \eq{SCETcurrentO2} we sum over directions $\vect{n}_i$ of the light-cone vectors $n_i = (1,\vect{n}_i)$. In the center-of-mass frame the two jet directions are constrained to be back-to-back, which eliminates the sum over one of the directions $\vect{n}_i$, leaving a sum over a single $\vect n$. 
The field redefinitions replace all usoft-collinear interactions with usoft Wilson lines, $Y_{n_i},\overline Y_{n_i}$, in the directions and color representations of the corresponding collinear fields.

Since collinear fields in different directions do not couple to one another in SCET at leading order, the forward matrix element in \eq{QCD2} vanishes, unless the directions $\vect n$ of the operator ${\cal O}$ and its complex conjugate agree with one another. By the same argument, the label momenta on these fields and their complex conjugates have to agree with one another. 
This gives
\begin{equation}
\begin{split}
\frac{d\sigma}{de} = & \frac{1}{2Q^2}\sum_{\vect{n}}\sum_{\tilde p_1, \tilde p_2}
C_{n\bar n}^*(\tilde p_1,\tilde p_2;\mu)C_{n\bar n}(\tilde p_1,\tilde p_2;\mu)\int\! d^4 x\,\e^{iQ x^0}\e^{i(\tilde p_2-\tilde p_1)\cdot x} \\
&\times\sum_{i=V,A}L^i_{\mu\nu}\bra{0}
\left[\bar\chi_{\bar n,p_2}^a\overline Y_{\bar n}^{\dag ab}\bar\Gamma^i_\mu Y_n^{\dag bc}\chi_{n,p_1}^c\right](x)
\delta(e -\hat e)
\left[\bar\chi_{n,p_1}^dY_n^{de}\Gamma^i_\nu\overline Y_{\bar n}^{ef}\chi_{\bar n,p_2}^f\right](0)\ket{0}\,,
\end{split}
\label{SCETdist}
\end{equation}
where we have made all color indices explicit. Finally, we demand that the label momenta that appear in the exponentials of this relation equal the total momentum, thus requiring $\bar n\cdot\tilde p_1 = -n\cdot\tilde p_2 = Q$ and $\tilde p_1^\perp = \tilde p_2^\perp = 0$.

\subsection{SCET scaling and the event shapes}
\label{scetscalesec}

For our reasoning below, it will be important to estimate the
contributions to event shapes from the label and residual momenta
of collinear particles in the final state as well as from usoft particles.
To be  concrete, we consider the angularities $e = \tau_a$, 
from \eq{eventshpdef1},
for which $f_a(\eta) = \e^{-\abs{\eta}(1-a)}$,
and follow the logic of \cite{Berger:2003iw}. 
The contribution of an individual particle to $\tau_a(X)$ is
\be
\label{taucontribution}
\frac{\abs{\vect{p}_T}}{Q}\e^{-\abs{\eta}(1-a)} = \frac{\abs{\vect{p}_T}}{Q}\min\left(\frac{E \mp p_\|}{E \pm p_\|}\right)^{\frac{1-a}{2}}\,.
\ee
For usoft particles we have $p_T \sim Q \lambda^2$ and the ratio in parentheses $R\sim 1$, while for collinear particles $p_T \sim Q\lambda$ and  $R\sim \lambda^2$. 
The nominal contribution of a 
usoft particle to $\tau_a$ is thus $\lambda^2$, while for a 
collinear particle it is $\lambda^{2-a}$.   
This is the case for any value of parameter $\lambda\ll 1$,
For $a<0$, the scaling behavior of the event shapes is
$\lambda^2$, independent of $a$, and dominated by the momenta of
usoft particles, independent of the collinear particles. For $0<a<1$, the event shapes scale as $\lambda^{2-a}$.  We will establish our results below for all angularities with $a<1$ and related event shapes, keeping all contributions to the event shape which are at least of order $\lambda^2$.

For $a\geq 1$, the contribution of ``ultra-collinear'' particles, for which $p_T\sim Q\lambda^2$ and $R\sim \lambda^4$ become important  \cite{Lee:2006nr}. Their contribution to the event shape is of the order $\lambda^{2(2-a)}$, which is at least as large as $\lambda^2$ for $a\geq 1$. Such particles are described by collinear modes in SCET with purely residual (zero label) transverse momenta. Our analysis below applies to jets whose typical constituents have  nonzero transverse label momenta, and so is appropriate only for $a<1$.

The subdominance of ultra-collinear contributions for $a<1$ corresponds to the observation that at a given $p_T$,
wide-angle emission with energy comparable to
$p_T$ contributes to the thrust and related
event shapes more than does collinear emission 
with energies much larger than $p_T$ \cite{Korchemsky:1999kt}
at the level of both perturbative and nonperturbative corrections.

\subsection{Factorizing the vacuum matrix element}

At this point, the field content in the forward matrix element is completely factorized. To complete the factorization of the matrix element in Eq.~(\ref{SCETdist}), we must in addition show that the operator $\delta(e-\hat e)$ also factorizes into collinear and soft parts. From \eqs{ehatdef}{ETfromT0i} we can see that the operator $\hat e$ can be written in terms of the energy-momentum tensor, 
as well as the operator $\hat t$, which determines the thrust axis. As we show in Sec.~\ref{sec:thrust}, the thrust axis is determined solely by the labels on the fields in the two-jet operator ${\cal O}_{n\bn}$, for which one simply finds that $\vect{t} = \vect{n}$. We will also use our observations in Sec.~\ref{scetscalesec} above, that the event shape values, and hence the action of the operator $\hat e$ on final state, are dominated by the contributions of usoft momenta, collinear label momenta $\bar n_i\mcdot\tilde p$ and $\tilde p_\perp$, and only one component $n_i\mcdot k$ of collinear residual
momenta, for particles in each collinear direction $n_i$.

The energy-momentum tensor, which is defined as
\be
T^{\mu\nu}\equiv \sum_i\frac{\partial{\mathcal{L}}}{\partial(\partial_\mu\phi_i)}\partial^\nu\phi_i - g^{\mu\nu}\mathcal{L}\,,
\ee
where the sum is over fields $\phi_i$ in the Lagrangian $\mathcal{L}$,
simplifies in SCET since after the BPS field redefinition the Lagrangian separates into 
\be
\label{decoupledLagrangian}
{\cal L} =  {\cal L}_{n} + {\cal L}_{\bn} + {\cal L}_{\rm us}\,.
\ee
As a result the energy-momentum tensor is a direct sum over contributions from fields in the different sectors. We must remember, of course, that \eq{decoupledLagrangian} holds only at leading order in $\lambda$ in SCET, and that there are power-suppressed terms in the SCET Lagrangian in which interactions between collinear and usoft fields do not decouple following the BPS field redefintion \cite{Bauer:2002uv,Chay:2002vy,Pirjol:2002km,Bauer:2003mga}.

Using the definition of the event shape operator given in Eq.~\eqref{ehatdef}, we find the important result in SCET
\be
\label{splite}
\hat e = \hat e_{n} + \hat e_{\bn} + \hat e_{\rm us} \,,
\ee
where
\be
\hat e_i = \frac{1}{Q} \int_{-\infty}^\infty \!\!d\eta \, f_e(\eta) \mathcal{E}_T^i(\eta)\,,
\ee
and $\mathcal{E}_T^i(\eta)$ is defined using \eq{ETfromT0i}, but using the energy-momentum tensor derived from ${\cal L}_i$. 
This means that we can write the delta function constraining the value of the event shape to its observed value as 
\be
\delta(e-\hat e) = \int\! de_{n}\,  \delta(e_{n} - \hat e_{n}) \int\! de_{\bn} \, \delta(e_{\bn} - \hat e_{\bn}) \int\! de_{\rm us}\, \delta(e_{\rm us} - \hat e_{\rm us}) \,
\delta(e-e_n-e_\bn-e_{\rm us})\,.
\label{efactorization}
\ee
Finally, we use the fact that the operators $\hat e_n$ ($\hat e_\bn$) are constructed only from collinear fields in the $n$ ($\bn$) direction and $\hat e_{\rm us}$ only from usoft fields. Thus 
\be
[\hat e_{n},\chi_\bn] = [\hat e_{\bn},\chi_n]  =  [\hat e_{n},Y] = [\hat e_{\bn},Y] = [\hat e_{\text{us}},\chi_n] =  [\hat e_{\text{us}},\chi_\bn]=   0\,.
\ee 
This enables us to rewrite \eq{SCETdist} in factorized form, separating the vacuum expectation values of mutually commuting fields,
\begin{align}
\label{factorizedform}
\frac{d\sigma}{de} &=  \frac{1}{6Q^2}\sum_{\vect{n}}\abs{C_{n\bar n}(Q,-Q;\mu)}^2\int d^4 x\int de_n\,de_{\bar n}\,de_s\,\delta(e-e_n-e_{\bar n}-e_s)  \\
&\quad\times\frac{1}{N_C^2}\Tr\bra{0}\chi_{n,Q}(x)_\beta\delta(e_n-\hat e_n)\bar\chi_{n,Q}(0)_\gamma\ket{0} \Tr \bra{0}\bar\chi_{\bar n,-Q}(x)_\alpha \delta(e_{\bar n} - \hat e_{\bar n})\chi_{\bar n,-Q}(0)_\delta\ket{0} \nn \\
&\quad\times\Tr\bra{0}\overline Y_{\bar n}^\dag(x)Y_n^\dag(x)\delta(e_s - \hat e_s)Y_n(0)\overline Y_{\bar n}(0)\ket{0}\sum_{i=V,A}L^i(\bar\Gamma_i^\mu)_{\alpha\beta}(\Gamma_{i\mu})_{\gamma\delta}\,, \nn
\end{align}
where $L^i = g^{\mu\nu}L^i_{\mu\nu}$, the traces are over colors, and we now make the spin indices explicit.

The collinear matrix elements define jet functions $\mathcal{J}_{n,\bar n}$ according to
\begin{subequations}
\label{jetfunctions}
\begin{align}
\frac{1}{N_C}\Tr\bra{0}\chi_{n,Q}(x)_\beta\delta(e_n  -  \hat e_n)\bar\chi_{n,Q}(0)_\gamma\ket{0} &\equiv\! \int\!\frac{dk^+\! dk^- \! d^2 k_\perp}{2(2\pi)^4}e^{-ik\cdot x}\mathcal{J}_n(e_n,k^+;\mu)\!\left(\!\frac{\nslash}{2}\!\right)_{\!\!\beta\gamma}\\
\frac{1}{N_C}\!\Tr \bra{0}\bar\chi_{\bar n,-Q}(x)_\alpha\delta(e_{\bar n} -  \hat e_{\bar n})\chi_{\bar n,-Q}(0)_\delta\ket{0} &\equiv\! \int\frac{dl^+ dl^- d^2 l_\perp}{2(2\pi)^4}e^{-il\cdot x}\mathcal{J}_{\bar n}(e_{\bar n},l^-;\mu)\left(\!\frac{\bnslash}{2}\!\right)_{\!\!\delta\alpha} 
\,,
\end{align}
\end{subequations}
while the usoft matrix element defines a soft function
\be
\label{softfunction1}
\frac{1}{N_C}\Tr\bra{0}\overline Y_{\bar n}(x)Y_n^\dag(x)\delta(e_s - \hat e_s)Y_n(0)\overline Y_{\bar n}(0)\ket{0} \equiv \int \frac{d^4 r}{(2\pi)^4}e^{-ir\cdot x} S(e_s,r;\mu)\,.
\ee
Furthermore, we can use that
$\mathcal{J}_n(e_n,k^+;\mu)$ depends only on single light-cone component $k^+ \equiv n\mcdot k$ of the residual momentum, as only $i n\cdot\partial$ appears in the $n$-collinear SCET Lagrangian~\cite{Bauer:2000yr} at leading-order in $\lambda$. Similarly $\mathcal{J}_{\bar n}(e_{\bar n},l^-;\mu)$ depends only on $l^-\equiv\bar n\mcdot l$. 
To express the cross section, Eq.\ (\ref{factorizedform}), in terms of the jet and soft functions directly,
we follow a variation of the reasoning of Ref.\ \cite{Fleming:2007qr} for massive
quarks.    

Given the discussion of Sec.~\ref{scetscalesec} above, the residual transverse ($k_\perp$ and ${l}_\perp$) and parallel-moving
($k^-$ and $l^+$) components of the jet functions enter the angularity event shapes only at nonleading power in $\lambda$,
as long as the parameter $a\le 1$.   This is the case for all values of $\lambda\ll 1$, and therefore holds
for perturbative, logarithmic corrections to the angularity cross sections, as well as in the nonperturbative region.
For the following discussion, we therefore define the jet event shape
operators $\hat{e}_n$ and $\hat{e}_{\bar{n}}$ to measure only the residual components $k^+$ and $l^-$  in the collinear sector.
Corrections to this approximation from extending integrations over residual momenta
to order $Q$ are perturbative and nonlogarithmic, and hence
can be absorbed into the matching coefficients $C_{n\bar{n}}$.   For simplicity, however, we do not change
our notation for the matching coefficients to reflect this here.

Because the jet functions are independent of the residual transverse momenta,
it is natural to change variables to their sum and difference:
$K_\perp \equiv k_\perp+l_\perp$,
$\kappa_\perp=(k_\perp - l_\perp)/2$.  The integral over the sum gives $(2\pi)^2\delta^2(x_\perp)$.  In this notation, the formula \eq{factorizedform} for the event shape distribution can now be written as
\begin{align}
\frac{d\sigma}{de} &= \frac{N_C}{6Q^2}L\sum_{\vect{n}}
\int d^2 \kappa_\perp\
\abs{C_{n\bar n}(Q,-Q;\mu)}^2 \int d^4 x\int de_n\,de_{\bar n}\,de_s\,\delta(e-e_n-e_{\bar n}-e_s) \nn \\
&\quad\times\delta\left(\frac{x^+}{2}\right)\delta\left(\frac{x^-}{2}\right)\delta^2(x_\perp) 
\int\frac{dk^+   dl^-}{4(2\pi)^4} \int \frac{d^4 r}{(2\pi)^4}
\e^{-i(r^+ + k^+) x^-/2 - i(r^- +l^-) x^+/2 -ir_\perp\cdot x_\perp} \nn \\
&\quad\times\mathcal{J}_n(e_n,k^+;\mu)\mathcal{J}_{\bar n}(e_{\bar n},l^-;\mu)S(e_s,r;\mu)\,, 
\label{factorizedformsimpler}
\end{align}
where the factor $L$ is defined
\begin{equation}
L \equiv L^V  \Tr\left(\frac{\nslash}{2}\gamma_\perp^\mu \frac{\bnslash}{2}\gamma^\perp_\mu\right) + L^A  \Tr\left(\frac{\nslash}{2}\gamma_\perp^\mu\gamma_5 \frac{\bnslash}{2}\gamma^\perp_\mu\gamma_5\right)\,.
\end{equation}
In Eq.~(\ref{factorizedformsimpler}) we have integrated over $k^-,l^+$, and $K_\perp$ to generate delta functions setting all components of $x$ to zero. This allows us to perform the integrals over $x$. 

In Eq.~(\ref{factorizedformsimpler}) there remains a sum over label directions $\vect n$ and an integral over the residual momenta $\kappa_\perp$,
which combined 
are simply an integral over total solid angle \cite{Fleming:2007qr},
\begin{equation}
\sum_{\vect n} d^2 \kappa_\perp =  \frac{Q^2}{4}d\Omega \,,
\end{equation}
where the overall factor arises from the magnitude of the label three-momentum of the jet in direction $\vect{n}$, which is $\abs{\vect{\tilde p}} = Q/2$.
We carry out the integral over solid angle to obtain our final result. To simplify our expression we define the Wilson coefficient $C_2(Q;\mu)\equiv C_{n\bn}(Q,-Q;\mu)$ since it is independent of $\vect n$. We now define
jet and soft functions integrated over all momenta:
\begin{subequations}
\begin{align}
J_n(e_n;\mu) &\equiv \int\frac{dk^+}{2\pi}\mathcal{J}_n(e_n,k^+;\mu) \\
J_{\bar n}(e_\bn;\mu) &\equiv \int\frac{dl^-}{2\pi}\mathcal{J}_\bn(e_\bn,l^-;\mu) \\
S(e_s;\mu) &\equiv \int \frac{d^4r}{(2\pi)^4} S(e_s,r;\mu) \,.
\end{align}
\end{subequations}
We note that the differential cross section can be expressed in terms of the total Born cross-section 
for $e^+e^-\rightarrow q\bar q$ 
\begin{equation}
\sigma_0 = \frac{4\pi\alpha^2 N_C}{3 Q^2}\sum_f \left[Q_f^2 - \frac{2Q^2 v_e v_f Q_f}{Q^2 - M_Z^2} + \frac{Q^4(v_e^2+a_e^2)(v_f^2+a_f^2)}{(Q^2-M_Z^2)^2}\right]\,,
\end{equation}
and we find as the final result for the differential event shape distribution:
\begin{equation}
\label{finalformula}
\frac{1}{\sigma_0}\frac{d\sigma}{de} = \abs{C_2(Q;\mu)}^2 \int de_n\,de_{\bar n}\,de_s\,\delta(e-e_n-e_{\bar n}-e_s) J_n(e_n;\mu) J_{\bar n}(e_{\bar n};\mu) S(e_s;\mu)\,.
\end{equation}
This is the main result of this paper,  in which the hard scattering function $H(\mu)$ of \eq{factorizationtheorem} is identified with the absolute square of an SCET matching coefficient, and the jet and soft functions have been given effective theory definitions. Corrections suppressed at least by a power of $\lambda$ to this formula enter from effects at subleading order in the SCET power expansion.

\subsection{Comment on final states in SCET}

In earlier studies of event shape distributions and two-jet cross sections in SCET \cite{Bauer:2002ie,Bauer:2003di,Trott:2006bk,Lee:2006nr,Fleming:2007qr}, the hadronic final states $X$ appearing in the cross-sections were not summed over before matching the full and effective theories. In these cases, the state $X$ was itself factored into separate collinear and soft parts
\begin{equation}
\ket{X} = \ket{X_n}\ket{X_{\bar n}}\ket{X_{\text{us}}}.
\label{factoredstates}
\end{equation}
In these studies the factorization theorem \eq{finalformula} is of the same form, but the jet function $\mathcal{J}_n$ is defined by
\begin{equation}
\label{reinsertedstates}
\sum_{X_n}\!\bra{0}\chi_{n,Q}(x)_\beta\ket{X_n}\!\bra{X_n}\bar\chi_{n,Q}(0)_\gamma\ket{0}\delta(e_n \! - \! e(X_n)) \equiv \int\!\!\frac{d^4 k}{(2\pi)^4}e^{-ik\cdot x}\mathcal{J}_n(e_n,k^+;\mu)\!\left(\!\frac{\nslash}{2}\!\right)_{\!\!\beta\gamma},\\
\end{equation}
and similarly for ${\cal J}_{\bar n}$ and $S$.  Such definitions, of course, can be derived
from the vacuum matrix elements in Eqs.~(\ref{jetfunctions}) and (\ref{softfunction1}),
if we assume that each set of $X_n,X_\bn,X_{\text{us}}$ form individually a complete set of states.
We note, however, that
the fields $\chi_{n,\bn}$ overlap with states ${X_{n,\bn}}$ only if these states are color triplet. 
Similarly, the states ${X_{\text{us}}}$ in general carry non-singlet color. 
At this level, discussions that begin with the assumption of \eq{factoredstates} yield results that are
equivalent to those of our reasoning.   
We have outlined the conceptual drawbacks of the use of this assumption
in the Introduction.
The advantages of 
avoiding the separation of states as in Eq.~(\ref{factoredstates}),
however,
are not only conceptual.   By defining the jet functions in terms of 
matrix elements as in Eqs.~(\ref{jetfunctions}) and (\ref{softfunction1}), we can estimate 
corrections to the factorized cross section.   These are due on the one
hand to the leading-$\lambda$ approximation in the equation
for the QCD Lagrange density, (\ref{decoupledLagrangian}), and on the other hand
to our ability to calculate or otherwise estimate through power corrections
\cite{Lee:2006nr} the specific vacuum matrix elements involving the operators $\hat{e}_i$.

We may also compare the role of partonic final states in SCET to their
treatment  in Ref.~\cite{Berger:2003iw},
in which factorized jet and soft functions
are each associated with individual sums over final states.
In Ref.~\cite{Berger:2003iw}, as in previous derivations
of resummed event shapes \cite{Catani:1992ua,Dokshitzer:1998kz}, final states
are always partonic.
 The factorization is carried out
at the level  of the diagrammatic cross section
order-by-order in the coupling and region-by-region
in phase space.   Each such region is characterized by
a definite set of jet and soft final-state partons, which are grouped
into factorized jet and soft functions.
In Ref.~\cite{Berger:2003iw}, each of these factors is initially
associated with a limited phase  space for its final-state partons.
These initial sums over partonic
final states are then extended into unconstrained
phase space sums for the jet and final states by
redefining the functions systematically to  avoid double counting
in any region that is infrared sensitive.    This
procedure relies on the exponentiation
properties of eikonal cross sections \cite{Berger:2002sv,Frenkel:1984pz,Gatheral:1983cz,Sterman:1981jc}.

In the diagrammatic treatment just described, specific classes of nonperturbative power 
corrections
may be inferred from ambiguities in the resummed expressions \cite{Beneke:2000kc}.
A full treatment of nonperturbative corrections
would require analyzing the observables in terms
of matrix elements of the full theory,  as advocated
in \cite{Sveshnikov:1995vi,Korchemsky:1997sy,Lee:2006nr},
a viewpoint that is consistent with the SCET analysis described above.

\section{Definition of required operators}
\label{sec:defofoperators}

In this section we explicitly construct the operator $\hat e$ defined in \eq{ehatdef}, which we used to eliminate the dependence on the final hadronic state ${X}$ of the weight $\delta(e-e(X))$ appearing in the event shape distribution $d\sigma/de$. This operator $\hat e$ itself depends on the transverse momentum flow operator $\mathcal{E}_T(\eta)$ and the thrust axis operator $\hat t$. We first confirm \eq{ETfromT0i} giving $\mathcal{E}_T(\eta)$ in terms of the energy-momentum tensor $T_{\mu\nu}$, and then argue that $\hat t$ can be replaced with an axis $\vect{t_L}$ that depends only on labels in the effective theory two-jet operator up to subleading power corrections (at least for a large set of event shapes which we identify).

\subsection{Energy Flow from Energy-Momentum Tensor}
\label{sec:EMtensor}

The transverse momentum flow operator $\mathcal{E}_T(\eta,\hat t)$ in Eq.~(\ref{ehatdef}) is related to the energy flow operator $\mathcal{E}(\hat n)$ defined in Ref.~\cite{Belitsky:2001ij}, whose action on states $\ket{X} = \ket{k_1,\dots,k_n}$ is
\begin{equation}
\mathcal{E}(\hat n)\ket{X} = \sum_{i\in X}\omega_i \delta^2(\hat n - \hat n_i)\ket{X},
\label{Endef}
\end{equation}
where $\omega_i$ is the energy of particle $i$, $\hat n$ is the unit vector pointing in the direction $(\theta,\phi)$, and $\delta^2(\hat n - \hat n_i) = \delta(\cos\theta-\cos\theta_i)\delta(\phi-\phi_i)$. To change variables from $\theta$ to $\eta$, we use the relation $\cos\theta = \tanh\eta$ for massless particles, to obtain
\be
\delta(\cos\theta - \cos\theta_i) = \delta(\tanh\eta - \tanh\eta_i) = \cosh^2\eta\, \delta(\eta-\eta_i)\,.
\ee
We define $\theta$ and $\eta$ to be measured with respect to the thrust axis. Also, the energy $\omega_i$ and the momentum $\abs{\vect{p}_i^T}$ transverse to the thrust axis are related by
\begin{equation}
\abs{\vect{p}_i^T} = \omega_i\sin\theta_i = \frac{\omega_i}{\cosh \eta_i }\,.
\end{equation}
Thus we can relate the transverse momentum flow operator in \eq{ETdef} to the energy flow operator in \eq{Endef} by
\begin{equation}
\mathcal{E}_T(\eta) = \frac{1}{\cosh^3\eta}\int_0^{2\pi}d\phi\,\mathcal{E}(\hat n).
\end{equation}
We will now verify that $\mathcal{E}(\hat n)$ is related to the energy-momentum tensor by
\ba
\mathcal{E}(\hat{n})
=
\lim_{R\to \infty}\, R^2\, \int_0^\infty dt\
\hat n_i\, T_{0i}\left(t,R\hat{n}\right )\,,
\label{Eflowdef}
\ea
the form of which was introduced in Ref.~\cite{Korchemsky:1997sy}.\footnote{A similar formula integrating over distance in the direction $\hat n$ with time taken to infinity is derived in Refs.~\cite{Cherzor:1997ak,Sveshnikov:1995vi}.} 
We will think of 
the variable $R$ as a truly macroscopic distance,
literally the distance from the scattering interaction
to a detector. Thus, $R$ is typically many orders of magnitude larger than the typical inverse minimum
mass scale of the process.

Consider the observable
\ba
F(\hat{n},q)
=
\sum_X \int d^4x \, e^{iq\cdot x}\bra{0} j^\dag(x) \mathcal{E}(\hat{n}) \ket{X} \bra{X}  j(0)\ket{0}\,,
\label{Fdef}
\ea
where the sum is over a complete set of out states $X$, and the current $j(x)$ couples to the field $\phi$.
Using the definition of the energy flow operators given in \eq{Endef}, one can easily show that this observable gives the weighted cross section 
\ba
F(\hat{n},q)
=
\sum_{\text{states }X}(2\pi)^4\, \delta^4\left(p_X - q\right)\!\!\!\!\!\!
\sum_{\text{particles } i \in X}\!\!\!\!\!\!\!\!\omega_i\, \delta^2\left(\hat{n} - \hat{n}_i\right) 
\bra{0} j^\dagger(0) \ket{X} \bra{X} j(0) \ket{0}\,,
\label{FomegaM}
\ea
where $\hat{n}_i$ is a unit vector in the direction of the three-momentum of particle $i$. 
Our aim is to show that the representation of the energy flow operator in terms of the energy-momentum tensor, as given in \eq{Eflowdef}, reproduces this result, with corrections entering with inverse powers of the
distance $R$ from the interaction region to the detector.

To begin, we insert another sum over out states
$X'$ between the current $j(x)$ and $\mathcal{E}$ in Eq.~(\ref{Fdef}).  This gives
\ba
F(\hat{n},q)
=
\sum_{\text{states }X, X'}
\int d^4x \, e^{iq\cdot x} \bra{0} j(x)\ket{X} 
\bra{X} \mathcal{E}(\hat{n})\ket{X'}
\bra{X'} j(0)\ket{0}\,.
\label{Ffree}
\ea
We then observe  that since
$\mathcal{E}(\hat n)$ is at the macroscopic distance $R$ from the scattering, all hadrons in the states $X,X'$
will have stopped interacting by the time they reach the position of
the operator $\mathcal{E}(\hat{n})$. The matrix elements of $\mathcal{E}(\hat n)$ between states $X,X'$ are thus those of a free hadronic theory.

To prove \eq{Eflowdef}, we must show that $\mathcal{E}(\hat n)$ acts on these hadronic states according to \eq{Endef}. We do so by plugging the energy-momentum tensor $T_{\mu\nu}$ appropriate for hadrons of a given type into \eq{Eflowdef} and testing its action on these states in the appropriate free field theory. This works for hadrons of any spin. Below we will demonstrate this explicitly for real scalars and Dirac fermions. Note that the scalar and Dirac fields represent the hadronic final states, not the partonic states. Since the hadrons are non-interacting, we only need to consider free field theories.

\subsubsection{Scalar fields}

We will first evaluate the matrix elements of $\mathcal{E}(\hat n)$ in \eq{Ffree} for a free, neutral scalar 
field, $\phi(x)$, for which
\ba
T_{0i}(x) =  \pi(x) \partial_i\, \phi(x)\, ,
\label{Tpiphi}
\ea
with $\pi(x) =\dot{\phi}(x)$ the corresponding conjugate momentum.

In the free scalar theory we can expand the energy-momentum tensor, and thus $\mathcal{E}(\hat n)$, 
in terms of particle creation and annihilation operators. 
Employing the mode expansion of $T_{0i}$ in \eq{Eflowdef},
we obtain for $\mathcal{E}(\hat n)$ in the free theory,
\begin{equation}
\begin{split}
\mathcal{E}(\hat n)
&= \lim_{R\rightarrow\infty} R^2\int_0^\infty\! dt\int\frac{d^3 p}{(2\pi)^3 2\omega_{\vect{p}}}\int\frac{d^3 q}{(2\pi)^3 2\omega_{\vect{q}}}\omega_{\vect{p}}\hat n\cdot\vect{q} \\
&\qquad\qquad\times\Bigl[a_{\vect{p}} a_{\vect{q}}e^{-i(\omega_{\vect{p}}+\omega_{\vect{q}})t}e^{iR\hat n\cdot(\vect{p}+\vect{q})} - a_{\vect{p}}a_{\vect{q}}^\dag e^{-i(\omega_{\vect{p}}-\omega_{\vect{q}})t}e^{iR\hat n\cdot(\vect{p}-\vect{q})} + \text{h.c.}\Bigr]\,.
\end{split}
\end{equation}
We have chosen a normalization for which $[a_\vect{p},a_{\vect{q}}^\dag] = 2\omega_{\vect{p}}(2\pi)^3\delta^3(\vect{p}-\vect{q})$.
We evaluate the integrals in spherical coordinates, performing the angular $\theta_{p,q},\phi_{p,q}$ integrals in the stationary phase approximation. We find the stationary point $\theta_{p,q} = \theta_{\hat n}$ and $\phi_{p,q} = \phi_{\hat n}$, that is, $\vect{p},\vect{q}$ are aligned with $\hat n$. Then,
\begin{align}
\label{stationaryintegral}
\mathcal{E}(\hat n)
= 
\lim_{R\rightarrow\infty}\frac{1}{4(2\pi)^4}\int_{-R}^\infty\! dx^+ &\int_0^\infty\! dp\,p\int_0^\infty\! dq\,q
\frac{q}{\omega_{\vect{q}}}\biggl[a_{p\hat n}a_{q\hat n}^\dag e^{-i(\omega_{\vect{p}}-\omega_{\vect{q}})x^+}e^{-iR(p^+ - q^+)} \\
&\qquad\qquad\qquad\qquad- a_{p\hat n}a_{q\hat n} e^{-i(\omega_{\vect{p}}+\omega_{\vect{q}})x^+}e^{-iR(p^+ + q^+)} +\text{h.c.}\biggr], \nn
\end{align}
where we made use of the light-cone coordinates with respect to $n = (1,\hat n)$, $x^+ = n\cdot x = t-R$, $p^+ = n\mcdot p =  \omega_{\vect{p}} - p$.
The $R\rightarrow\infty$ limit of the $x^+$ integral produces delta functions $(2\pi)\delta(\omega_{\vect{p}}-\omega_{\vect{q}})$ and $(2\pi)\delta(\omega_{\vect{p}}+\omega_{\vect{q}})$. The latter delta function cannot be satisfied, so we are left with simply
\begin{equation}
\mathcal{E}(\hat n)
 = \int\frac{d^3 p}{(2\pi)^3 2\omega_{\vect{p}}} \omega_{\vect{p}} \,a_{\vect{p}}^\dag a_{\vect{p}} \delta^2(\hat n - \hat p),
\label{Enfreefinal}
\end{equation}
where we have normal-ordered the operators in Eq.~(\ref{stationaryintegral}) and dropped the infinite constant term. The operator thus picks out the total energy of all scalar hadrons in the direction $\hat n$. Inserting the result into Eq.~(\ref{Ffree}) we reproduce the weighted cross section given in \eq{FomegaM}, completing the proof.

There are corrections to this result from terms dropped in the stationary phase approximation used in Eq.~(\ref{stationaryintegral}) of order $\mathcal{O}(1/R)$, or, more precisely of order $\mathcal{O}(1/(m_0 R))$, where $m_0$ is the smallest momentum scale that occurs naturally in the $S$-matrix elements. Although possibly small, it is still the inverse of some microscopic length scale. When multiplied by the macroscopic scale $R$, the result is a very large dimensionless number such that these corrections can safely be neglected.

\subsubsection{Dirac Fermions}

We can repeat the above derivation for Dirac fermions. The canonical energy-momentum tensor is
\begin{equation}
T_{\mu\nu} = i\bar\psi\gamma_\mu\partial_\nu\psi - g_{\mu\nu}\mathcal{L}.
\label{canonicalDiracT}
\end{equation}
Using the mode expansion of $T_{0i}$
 in Eq.~(\ref{Eflowdef}) gives for $\mathcal{E}(\hat n)$
\begin{align}
\mathcal{E}(\hat n) &= \lim_{R\rightarrow\infty}R^2\int_0^\infty\! dt\int\!\frac{d^3 p}{(2\pi)^3 2\omega_{\vect{p}}}\int\!\frac{d^3 q}{(2\pi)^3 2\omega_{\vect{q}}}\hat n\!\cdot\!\vect{q} \\
&\quad\times\sum_{r,s}\Bigl[b_{\vect{p}}^r b_{\vect{q}}^{s\dag} v^{r\dag}(p) v^s(q)\e^{-i(\omega_{\vect{p}}-\omega_{\vect{q}})t}\e^{iR\hat n\cdot(\vect{p}-\vect{q})} 
- a_{\vect{p}}^{r\dag}a_{\vect{q}}^s u^{r\dag}(p)u^s(q)\e^{i(\omega_{\vect{p}}-\omega_{\vect{q}}) t}\e^{-iR\hat n\cdot(\vect{p}-\vect{q})} \nn \\
&\qquad\quad -b_{\vect{p}}^{r}a_{\vect{q}}^s v^{r\dag}(p)u^s(q)\e^{-i(\omega_{\vect{p}}+\omega_{\vect{q}}) t}\e^{iR\hat n\cdot(\vect{p}+\vect{q})} +a_{\vect{p}}^{r\dag}b_{\vect{q}}^{s\dag} u^{r\dag}(p)v^s(q)\e^{i(\omega_{\vect{p}}+\omega_{\vect{q}}) t}\e^{-iR\hat n\cdot(\vect{p}+\vect{q})}\Bigr]. \nn
\end{align}
Performing the angular integrals in the stationary phase approximation and taking the $R\rightarrow\infty$ limit of the $t$ (or $x^+ = t - R$) integral as in the previous section, we are left with the terms
\begin{equation}
\mathcal{E}(\hat n ) = \frac{1}{4(2\pi)^3}\int_0^\infty dp\,p\frac{p}{\omega_{\vect{p}}}\bigl[a_{p\hat n}^\dag a_{p\hat n} u^{r\dag}(p\hat n) u^s(p\hat n) - b_{p\hat n}^r b_{p\hat n}^{s\dag}v^{r\dag}(p\hat n)v^s(p\hat n)\bigr].
\end{equation}
The spinors satisfy the relations
\begin{equation}
u^{r\dag}(p\hat n)u^s(p\hat n) = 2\omega_{\vect{p}}\delta^{rs},\quad v^{r\dag}(p\hat n)v^s(p\hat n) = 2\omega_{\vect{p}}\delta^{rs},
\end{equation}
and we obtain the final form of $\mathcal{E}(\hat n)$,
\begin{equation}
\mathcal{E}(\hat n) = \int\frac{d^3 p}{(2\pi)^3 2\omega_{\vect{p}}} \omega_{\vect{p}}\sum_s(a_{\vect{p}}^{s\dag} a_{\vect{p}}^s + b_{\vect{p}}^{s\dag} b_{\vect p}^s)\delta^2(\hat p-\hat n), 
\end{equation} 
after normal-ordering the $bb^\dag$ term and dropping the infinite constant. In this form, $\mathcal{E}(\hat n)$ picks out the energy of hadronic spin-$1/2$ Dirac fermions in direction $\hat n$.

\subsection{Defining the thrust axis}
\label{sec:thrust}

The factorization proof in Sec.~\ref{sec:twojet} makes use of an operator $\hat t$ which gives the thrust axis $\vect{t}(X)$ when acting on a final state ${X}$.
We begin by illustrating one explicit construction of $\hat t$ which we can use in full QCD. Then in SCET we argue that we can eliminate the use of the operator $\hat t$ entirely by identifying the thrust axis with the jet direction $\vect{n}$ in the two-jet current; an identification that is valid up to subleading power corrections in the event shapes we consider here.

\subsubsection{Thrust axis in QCD}
\label{sec:thrustQCD}

The thrust axis is the axis that is picked out by the maximum operation in the definition of thrust:
\begin{equation}
\label{thrustdef}
T= \frac{1}{Q}\max_{\vect{t}}\sum_{i}\abs{\vect{p}_i\cdot \vect{t}}\,.
\end{equation}
Each choice of axis $\vect{t}$ divides space into two hemispheres, $A,B$, so that the sum in \eq{thrustdef} splits into two corresponding pieces,
\begin{equation}
T = 
\frac{1}{Q}\max_{\vect{t}}\,(\vect{p}_A - \vect{p}_B)\cdot\vect{t}\,,
\end{equation}
where $\vect{p}_{A,B}$ are the total three-momentum in hemisphere $A,B$. By momentum conservation, $\vect{p}_A = -\vect{p}_B$, so 
\begin{equation}
\label{thrustA}
T = \max_{\vect{t}}\frac{2\vect{p}_A\cdot\vect{t}}{Q}\,,
\end{equation}
which is maximized by choosing
\begin{equation}
\vect{t} = \frac{\vect{p}_A}{\abs{\vect{p}_A}}\,,
\end{equation}
the unit vector in the direction of the total three-momentum in hemisphere $A$. Thrust then simplifies to
\begin{equation}
\label{thrustsimple}
T = \max_A\frac{2\abs{\vect{p}_A}}{Q}\,.
\end{equation}
Thus, to find the thrust, we need simply find the hemisphere with the largest total three-momentum, and the direction of this three-momentum is the thrust axis. To construct the operator which gives this axis, define the 4-vector of momentum flow operators,
\begin{equation}
\mathcal{P}_\mu(\hat n) \equiv \bigl(\mathcal{E}(\hat n),\vect{P}(\hat n)\bigr )= \lim_{R\rightarrow\infty}R^2\int_0^\infty dt\,\hat n_i T_{\mu i}(t,R\hat n)\,,
\end{equation}
extending the definition of the energy flow operator $\mathcal{E}(\hat n)$. 
Then the thrust axis operator is
\begin{equation}
\hat t = \mathcal{N}\left[\max_{A} \int_A d\Omega\,\vect{P}(\hat n)\right]\,,
\end{equation}
where the quantity maximized by the $\max$ operator is  the length of its three-vector argument, the integral is over the hemisphere $A$, and the operator $\mathcal{N}$ normalizes three-vectors, $\mathcal{N}(\vect{v}) = \vect{v}/\abs{\vect{v}}$. This construction manifests that the thrust axis of an event is just a function of particles' three-momenta and can be written in terms of the energy-momentum tensor in similar manner to the event shapes themselves.

\subsubsection{Thrust axis in SCET}
\label{sec:thrustSCET}

In SCET, we can replace $\hat t$ with the value $\vect{t_L}$ of a ``label'' thrust axis whose determination is very simple. Namely, $\vect{t_L}$ is determined by the momentum labels in the two-jet operator $\mathcal{O}_{n_1 n_2}$ and returns the thrust axis $\vect{\tL}(\tilde p_1,\tilde p_2)$ determined by these label momenta. The axis $\vect{t_L}$ thus depends not on the final state $X$ but only on the operator $\mathcal{O}_{n_1 n_2}$. For back-to-back jets, this ``label'' thrust axis is simply in the direction of the vector $\vect{n}$ in the operator $\mathcal{O}_{n\bar n}$. In this section we argue that identifying this label thrust axis with the true thrust axis is valid up to power corrections that are subleading in the SCET expansion parameter $\lambda$ for angularities with $a < 1$. For $a\ge 1$ these corrections become leading order, and the thrust axis cannot be determined from label momenta alone.

The true thrust axis $\vect{t}$ is that which maximizes the sum
\begin{equation}
T = \frac{1}{Q}\sum_i\abs{\vect{t}\cdot\vect{p}_i}\,,
\label{fullsum}
\end{equation}
where $\vect{p}_i$ are the full three-momenta of all the particles in the event. The ``label'' thrust axis $\vect{\tL}$ is defined to maximize
\begin{equation}
T_L = \frac{1}{Q}\sum_i\abs{\vect{\tL}\cdot\vect{\tilde p}_i}\,,
\label{labelsum}
\end{equation}
where $\vect{\tilde p}_i$ are the label momenta of all the particles in the event. 
The terms in \eq{labelsum} corresponding to each of the two jets may be grouped together, so that the individual label momenta in each group sum up to equal the total label momentum of each jet. This total label momentum is given by the label on the corresponding collinear field in the operator $\mathcal{O}_{n_1n_2}$, so $\vect{t_L}$ depends not on the final state but only on this operator. In SCET, then, each of the operators $\hat e_n,\hat e_\bn,\hat e_{\text{us}}$ in \eq{splite} depends on this axis $\vect{t_L}$, which in turn is completely determined by the choice of operator $\mathcal{O}_{n_1 n_2}$.

As we will show, the error induced in the event shape distribution by approximating the true thrust axis $\vect{t}$ by the label thrust axis $\vect{\tL}$ will be a subleading correction in the SCET expansion parameter $\lambda$. The only condition is that  the event shape in question sufficiently suppresses the contribution of collinear particles close to the thrust axis, as is the case for angularity distributions with $a<1$. To prove this, we will show that the two thrust axes are related by transformations of the light-cone vectors $n_i$ which leave the effective theory Lagrangian invariant, a property known as reparametrization invariance (RPI) \cite{Manohar:2002fd}. They induce variations of the operators $\mathcal{O}_{n\bar n}$ and $\hat e$ appearing in $d\sigma/de$ in \eq{SCETdist} only at subleading order in $\lambda$.

The true and label thrust axes differ by a quantity $\delta\vect{t}$, where $\vect{t} = \vect{\tL} + \delta\vect{t}$. The expression \eq{fullsum} that must be maximized to determine the true $\vect{t}$ is
\begin{equation}
\frac{1}{Q}\sum_{i}\abs{\vect{t}\cdot(\vect{\tilde p_i} + \vect{k}_i)},
\end{equation}
where we expanded the momentum $\vect{p}_i$ into its label and residual parts (for soft particles $\tilde p_i = 0$). This expression is identical, through terms of order $\lambda$, to the expression that must be maximized to determine the label thrust axis $\vect{\tL}$, \eq{labelsum}. Thus $\vect{t}=\vect{\tL}$ through order $\lambda$, and we conclude that $\delta\vect{t}$ is of order $\lambda^2$.
Furthermore, since $\vect{t}$ and $\vect{\tL}$ are both unit vectors, we have
\be
1 = \abs{\vect{t}}^2 = \abs{\vect{\tL} + \delta\vect{t}}^2 = \abs{\vect{\tL}}^2 + 2\vect{\tL}\cdot\delta\vect{t} + \abs{\delta\vect{t}}^2 = 1+ 2\vect{\tL}\cdot\delta\vect{t} + \mathcal{O}(\lambda^4)\,.
\ee
Thus $\vect{\tL}\cdot\delta\vect{t} = 0 + \mathcal{O}(\lambda^4)$; that is, $\delta\vect{t}$ is orthogonal to $\vect{\tL}$ up to terms of order $\lambda^4$. For a two-jet operator $\mathcal{O}_{n\bn}$, the label thrust axis $\vect{\tL}$ is just $\vect{n}$, and $\delta\vect{t}$ is purely transverse to $\vect{n}$: $\delta\vect{t} = \delta\vect{t}_\perp + \mathcal{O}(\lambda^4)$. 

Due to RPI, each of the collinear sectors in SCET is invariant under changes in the light-cone vectors $n_i$ and $\bn_i$ that leave the conditions $n_i^2 = \bn_i^2 = 0$ and $n_i\cdot\bn_i = 2$ unchanged. Here, we are considering an arbitrary number of collinear sectors each labeled by $i$, and associated with a corresponding light-cone direction $n_i$, where $n^\mu_i = (1,\vect{n}_i), \bn_i^\mu = (1,-\vect{n}_i)$ with $\vect{n}^2_i = 1$. Note that for only two label directions, momentum conservation fixes $n^\mu_1 = \bn^\mu_2 \equiv n^\mu$ and $\bn^\mu_1 = n^\mu_2 \equiv \bn^\mu$, however, for the analysis below we will continue to distinguish $i=1,2$ until the very end.

Under RPI type-I and type-II transformations \cite{Manohar:2002fd} in the $i$th collinear sector the four-vectors $n_i,\bn_i$ are shifted by a transverse pieces $\Delta^\perp_i,\varepsilon^\perp_i$,
\be
\label{RPI12}
\text{(I)}\  n_i \rightarrow n_i + \Delta^\perp_i \qquad \text{(II)}\ \bn_i \rightarrow \bn_i + \varepsilon^\perp_i\,,
\ee
where $\Delta^\perp_i$ is allowed to be $\mathcal{O}(\lambda)$ or smaller, and $\varepsilon^\perp_i$ is allowed to be $\mathcal{O}(1)$ or smaller. Below we will choose both $\Delta^\perp_i,\varepsilon^\perp_i$ to be only $\mathcal{O}(\lambda^2)$.
Under these transformations label momenta $\tilde p^\mu = \bn_i\mcdot\tilde p\,n^\mu/2 + \tilde p_{\perp }^\mu$ in the $i$th sector transform as 
\begin{subequations}
\label{RPIlabels}
\begin{align}
\text{(I)}\ \tilde p^\mu &\to \tilde {p}'^\mu  = \tilde p^\mu - \frac{1}{2} \Delta_i^\perp\mcdot\tilde p^\perp \bar{n}_i^\mu \\
\text{(II)}\ \tilde p^\mu &\to \tilde {p}'^\mu  = \tilde p^\mu - \frac{1}{2} \varepsilon_i^\perp\mcdot\tilde p^\perp {n}_i^\mu\,.
\end{align}
\end{subequations}

Since the label thrust axis $\vect{\tL}$ depends on label momenta in each collinear sector,  $\vect{\tL}$ also transforms under reparameterizations of $n_i,\bn_i$ in any sector: $\vect{\tL} \to \vect{\tL}' = \vect{\tL} + \delta\vect{\tL^i}$. Thus we can bring $\vect{\tL}$ to coincide with the full thrust axis $\vect{t}$ by performing a suitable set of RPI transformations on $n_i,\bn_i$. For the case of two jets, we will use an RPI type-I and a type-II transformation of each of $n_1$ and $n_2$ to  bring $\vect{\tL}$ to equal $\vect{t}$. To find the correct transformations, we require that the full thrust axis $\vect{t}$ maximize the sum in \eq{labelsum} after the label momenta are transformed as in \eq{RPIlabels}. After these transformations, the sum takes the form
\be
\label{transformedlabelsum}
\begin{split}
\sum_{j}\abs{\vect{x}\cdot\vect{\tilde p}'_j} &= \sum_{j\in 1}\abs{\vect{x}\cdot\left[\vect{\tilde p}_j - \frac{1}{2}(\vect{\Delta}_1^\perp\cdot\vect{\tilde p}_j^\perp)\vect{n}_1 +  \frac{1}{2}(\boldeps_1^\perp\cdot\vect{\tilde p}_j^\perp)\vect{n}_1\right]}  \\
&+ \sum_{j\in 2}\abs{\vect{x}\cdot\left[\vect{\tilde p}_j - \frac{1}{2}(\vect{\Delta}_2^\perp\cdot\vect{\tilde p}_j^\perp)\vect{n}_2 +  \frac{1}{2}(\boldeps_2^\perp\cdot\vect{\tilde p}_j^\perp)\vect{n}_2\right]}\,,
\end{split}
\ee
where the sum is divided into the $n_1$ and $n_2$ sectors, and $\vect{x}$ is the variable vector that must be chosen to maximize the sum. Let us use that $\vect{n}_1 = -\vect{n}_2 \equiv\vect{n}$, and consider RPI transformations such that $\vect{\Delta}_1^\perp = -\vect{\Delta}_2^\perp\equiv \vect{\Delta}_\perp$ and $\boldeps_1^\perp= -\boldeps_2^\perp \equiv\boldeps_\perp$. Then  we can write \eq{transformedlabelsum} as
\be
\label{xsum}
\sum_{j}\abs{\biggl[\vect{x} - \frac{1}{2}(\vect{x}\cdot\vect{n})(\vect{\Delta}_\perp - \boldeps_\perp) \biggr]\cdot\vect{\tilde p}_j}\,.
\ee
But this is the same form as the sum giving the label thrust axis in \eq{labelsum}, so we know that \eq{xsum} is maximized by choosing $\vect{x}$ such that the vector in brackets is just  $\vect{\tL}$ itself,
\be
\vect{x} - \frac{1}{2}(\vect{x}\cdot\vect{n})(\vect{\Delta}_\perp-\boldeps_\perp) = \vect{\tL}\,.
\ee
Using our choice of power counting, $\Delta_\perp,\varepsilon_\perp\sim\mathcal{O}(\lambda^2)$, we can solve for $\vect{x}$ order-by-order in $\lambda$. The solution starts as $\vect{x}= \vect{\tL} + \mathcal{O}(\lambda^2)$, and we can use $\vect{\tL}\cdot\vect{n} = 1$, to obtain
\be
\vect{x} = \vect{\tL} + \frac{1}{2}(\vect{\Delta}_\perp - \boldeps_\perp)\,.
\ee
This new solution for the label thrust axis coincides with the true thrust axis if we choose 
\be
\label{correctRPI}
\vect{\Delta}_\perp = -\boldeps_\perp = \delta\vect{t}\,.
\ee
In terms of four-vectors, \eq{correctRPI} corresponds to the set of RPI transformations,
\be
\begin{split}
n_1 \rightarrow (1,\vect{n}_1+\delta\vect{t})\,,&\quad \bn_1\rightarrow (1,-\vect{n}_1-\delta\vect{t}) \\
n_2\rightarrow (1,\vect{n}_2-\delta\vect{t})\,,&\quad \bn_2\rightarrow (1,-\vect{n}_2+\delta\vect{t})\,.
\end{split}
\ee
The label thrust axis corresponding to the operator $\mathcal{O}_{n_1n_2}$ is thereby brought to coincide with the true thrust axis. These leave the SCET Lagrangian invariant. The operator $\mathcal{O}_{n\bar n}$ is invariant up to corrections subleading in $\lambda$.  

Finally, we need to estimate the size of the variation in the operator $\hat e$ given in \eq{ehatdef}, but with the full thrust axis replaced by the label thrust axis, which is just the direction $\vect{n}$ appearing in $\mathcal{O}_{n\bn}$. Consider again the example of angularities $e = \tau_a$, for which $f_a(\eta) = \e^{-\abs{\eta}(1-a)}$, following \cite{Berger:2003iw}. As in \eq{taucontribution}, the contribution of an individual particle to $\tau_a(X)$ is
\be
\frac{\abs{\vect{p}_T}}{Q}\e^{-\abs{\eta}(1-a)} = \frac{\abs{\vect{p}_T}}{Q}\min\left(\frac{E \mp p_\|}{E \pm p_\|}\right)^{\frac{1-a}{2}}\,.
\ee
We recall from Sec.~\ref{scetscalesec} that for usoft particles $p_T \sim Q \lambda^2$ and the ratio in parentheses $R\sim 1$, while for collinear particles $p_T \sim Q\lambda$ and  $R\sim \lambda^2$. The shift in the thrust axis identified above induces for usoft particles $\delta p_T \sim Q \lambda^4$ and $\delta R\sim \lambda^2$, while for collinear particles we find $\delta p_T \sim Q \lambda^2$ and $\delta R\sim \lambda^3$. This changes the total contribution of a usoft particle to $e(X)$ by  $\delta e_s\sim\lambda^4$, and of a collinear particle by $\delta e_c\sim\lambda^{3-a}$. 

As long as $a<1$, $\delta e_c$ is smaller than $\lambda^2$. If, as in this paper, we are interested in calculating the event shape accurately only to order $\lambda^2$, then we may neglect the shifts $\delta e_s$ and $\delta e_c$. Beginning with $a=1$, for which $\tau_1 = B$ (broadening), the power corrections induced by the shift in the thrust axis cannot be neglected, since $\delta e_c \sim \lambda^{3-a}$ becomes as large as the terms we considered in this paper. The necessity of accounting for this ``recoil'' of the thrust axis against usoft-scale momenta was demonstrated in full QCD for $B$ in \cite{Dokshitzer:1998kz} and for $\tau_a$ with $a\geq 1$ in \cite{Berger:2003iw}.

\section{Conclusions}
\label{sec:conclusions}

We have proved the factorization of two-jet event shape distributions in soft collinear effective theory without assuming the factorization of hadronic final states into separate colored collinear and soft sectors as in previous discussions. To do this we expressed the weight of each final state $\delta(e-e(X))$ in the differential cross section $d\sigma/de$ as an operator built out of energy flow and thrust axis operators acting on the final state, allowing us to sum over a complete set of hadronic states before factorizing soft and collinear matrix elements. These results are valid up to $\mathcal{O}(\lambda)$ corrections to the decoupling of usoft and collinear degrees of freedom in SCET, $\mathcal{O}(1/R)$ corrections to the relation between the energy flow operator and energy-momentum tensor, and $\mathcal{O}(\lambda^b)$ corrections due to the difference between  the thrust axis and the collinear jet direction $\vect{n}$, where the power $b$ depends on the event shape in question.  Similar methods should also be useful in studying the factorization of other jet observables in both leptonic and hadronic collisions.

\acknowledgments
We thank I. Stewart for valuable discussions. CL is grateful to the C.N. Yang Institute for Theoretical Physics for its hospitality during a portion of this work. The work of CWB and CL was supported in part by the Director, Office of Science, Offices of High Energy and Nuclear Physics of the U.S. Department of Energy under Contract DE-AC02-05CH11231, and in part by the National Science Foundation under grant  PHY-0457315. The work of CL was also supported in part by the U.S. Department of Energy under Contract 
DE-FG02-00ER41132. The work of SF was supported in part by the U.S. Department of Energy under Contract DE-FG02-06ER41449. CWB would also like to acknowledge support from the
DOE OJI program, and an LDRD from LBNL.
The work of GS was supported in part by the National Science Foundation, 
grants PHY-0354776, PHY-0354822 and PHY-0653342.

\bibliography{EventShp}

\end{document}